\documentclass[twocolumn]{aastex62}

\usepackage{epsf}
\usepackage{amsmath}                
\usepackage{amsfonts}               
\usepackage{amssymb}                
\usepackage{epsfig}                 
\usepackage{float}
\usepackage{color}
\usepackage{tabularx}
\usepackage{comment}
\usepackage{makecell}

\def \eV{~\rm{eV}}

\def \s{~\rm{s}}
\def \km{~\rm{km}}

\def \K{~\rm{K}}

\def \g{~\rm{g}}
\def \G{~\rm{G}}

\def \erg{~\rm{erg}}

\def \yr{~\rm{yr}}

\begin{document}

\title{Mergers of neutron stars and black holes with cores of giant stars: a population synthesis study}


\author{Aldana Grichener}
\affiliation{Department of Physics, Technion, Haifa, 3200003, Israel; aldanag@campus.technion.ac.il}

\begin{abstract}
We perform population synthesis of massive binaries to study the mergers of neutron stars (NSs) and black holes (BHs) with the cores of their giant secondaries during common envelope evolution (CEE). We use different values of the efficiency parameter $\alpha_{\rm CE}$ in the framework of the energy formalism for traditional CEE ($\alpha_{\rm CE} \leq 1$) and including additional energy sources to unbind the envelope ($\alpha_{\rm CE} > 1$). We constrain the possible values of $\alpha_{\rm CE}$ by comparing the results of our simulations with local rate densities of binary compact object mergers as inferred from gravitational-wave observations. We find two main evolutionary pathways of binary systems that result in NS-core mergers, while only one of them can also lead to the merger of a BH with the core of the giant star. We explore the zero age main sequence (ZAMS) statistical properties of systems that result in NS/BH-core mergers and find that the two evolutionary channels correspond to a bimodal distribution of orbital separations. We estimate the percentage of the mergers' event rates relative to core collapse supernovae (CCSNe). We include the effect of mass accreted by the NS/BH during CEE in a separate set of simulations and find it does not affect the mergers' event rates.

\end{abstract}

\keywords{binaries: general -- stars: neutron stars -- stars: black holes  -- stars: massive   --methods: numerical}

\section{INTRODUCTION}
\label{sec:intro}

Most massive stars are in close multiple systems consisting of two or more stars in orbit around their common center of mass (e.g., \citealt{Sanaetal2012}). The swelling of one or both stars in a binary system at late evolutionary phases might lead to the filling of the Roche lobe around them to the point where mass transfer takes place. If the mass transfer becomes unstable then the system might enter a common envelope evolution (CEE) phase (e.g., \citealt{Paczynski1976}; \citealt{IbenLivio1993}; \citealt{TaamSandquist2000}; \citealt{Izzardetal2012}; \citealt{Ivanovaetal2013}; \citealt{RoepkeDeMarco2022}) in which the envelopes of both stars merge, or a stellar remnant with no envelope is immersed within the envelope of its companion. In cases where at the onset of CEE the stars in the binary are a neutron star (NS) or a black hole (BH) and a massive giant, the compact object \footnote{ Throughout the manuscript ``compact object'' refers only to a NS or a BH.} is swallowed by the giant star and spirals closer to its core. This can lead to either the ejection of the envelope and a surviving compact object-core pair or to the merger of the core with the NS/BH. 

Mergers of NSs/BHs with cores of giant stars are a topic of ongoing research in a variety of contexts (e.g., \citealt{FryerWoosley1998}; \citealt{ZhangFryer2001}; \citealt{BarkovKomissarov2011}; \citealt{Chevalier2012}; \citealt{Schroderetal2020}; \citealt{Metzger2022}; \citealt{Guarinietal2022}). In particular, such mergers can lead to transient events known as common envelope jet supernovae (CEJSNe; \citealt{SokerGilkis2018}). In a CEJSN event a NS/BH is engulfed by a giant star and spirals-in inside its envelope. The compact object accretes mass via an accretion disk and launches two opposite jets that propagate through the giant's envelope and expel mass. Eventually, the NS/BH reaches the dense core of the giant star and launches more energetic jets as they merge. The effects of the jets that the compact object launches on the envelope and on the core of the giant star are broadly studied in one dimensional and three dimensional simulations (e.g., \citealt{MorenoMendezetal2017}; \citealt{Moriya2018}; \citealt{Gilkisetal2019}; \citealt{LopezCamaraetal2019}; \citealt{Gricheneretal2021}; \citealt{Ragoleretal2022}; \citealt{Hilleletal2022}; \citealt{Schreieretal2022}). Another proposed alternative is the formation of a stable Thorne-$\rm \dot{Z}$ytkow object when a NS merges with the core of a red supergiant  (e.g., \citealt{ThorneZytkow1975}; \citealt{ThorneZytkow1977}; \citealt{Eichetal1989}).  

Unfortunately, the high opacity of the giant's envelope does not allow for direct observations of core-NS/BH mergers with existing facilities. Moreover, even though the merger of a compact object with the core of a giant star emits gravitational waves (e.g., \citealt{Ginatetal2020}), their signal is much lower than in the case of binary compact object mergers, and is undetectable at the moment. A future detection by next-generation gravitational waves detectors, such as LISA/DECIGO/BBO, can serve as an observational signature of a merger event. 

Meanwhile, we can study indirect evidence for NS/BH-core mergers. Due to the high energies of the jets in their correlated CEJSN transient, NS/BH-core mergers might account for several astrophysical phenomena, such as heavy r-process nucleosynthesis that occurs inside jets that a NS launches as it merges with the core of a giant star (\citealt{Papishetal2015}; \citealt{GrichenerSoker2019a}; \citealt{Gricheneretal2022}; \citealt{GrichenerSoker2022}) and high-energy neutrino emission from jets that a BH launches while spiraling-in inside the envelope of a giant prior to the merger with its core \citep{GrichenerSoker2021}. Moreover, some peculiar transient events such as fast blue optical transients and other rare supernovae (SNe) might hint to NS/BH-core mergers and the CEJSN mechanism as well (e.g., \citealt{Thoneetal2011}; \citealt{SokerGilkis2018}; \citealt{Sokeretal2019}; \citealt{Dongetal2021}; \citealt{Soker2022a}; \citealt{Soker2022b}). The recently proposed hypernebula transient that is powered by jets accompanying hyper-Eddington mass transfer from an evolved post-main sequence (MS) star onto a NS/BH shortly before CEE (\citealt{SridharMetzger2022}; \citealt{Sridharetal2022}) might serve as a precursor of a NS/BH-core merger. Estimating the mergers' event rates is crucial to understand whether they can account for said phenomena.

In this study we use the population synthesis code \textsc{COMPAS} \citep{Rileyetal2022} to find the merger rates of NSs/BHs with cores of giant stars. Many population synthesis studies of massive binaries were performed using codes such as \textsc{COMPAS} (e.g., \citealt{Stevensonetal2017}; \citealt{VignaGomezetal2018}; \citealt{Neijsseletal2019}; \citealt{VignaGomezetal2020}; \citealt{Broekgaardenetal2021}; \citealt{Ravehetal2022}; \citealt{Raufetal2023}), \textsc{BSE} (e.g., \citealt{Hurleyetal2002}; \citealt{Zhangetal2005}; \citealt{Hernandezetal2013}; \citealt{Shaoetal2021}), \textsc{\rm BINARY\textunderscore C} (e.g., \citealt{Izzardetal2006}; \citealt{Izzardetal2018}), \textsc{BPASS} (e.g., \citealt{Eldridgeetal2017}; \citealt{StanwayEldridge2018}; \citealt{StanwayEldridge2019}; \citealt{StevanceEldridge2021}; \citealt{Aadlandetal2022}; \citealt{Ghodlaetal2022}; \citealt{Pattonetal2022}), \textsc{STARTRACK} (e.g., \citealt{Belczynskietal2002}; \citealt{Belczynskietal2008}; \citealt{Oslowskietal2011}; \citealt{Dominiketal2012}; \citealt{Belczynskietal2016}; \citealt{Mondaletal2020}; \citealt{Perigoisetal2021}) and \textsc{SEVN}(e.g., \citealt{SperaMapelli2017}; \citealt{Speraetal2019}; \citealt{Mapellietal2020}; \citealt{Iorioetal2022}). To our knowledge, the only study which performed population synthesis of NS/BH-core mergers is \cite{Schroderetal2020} (see section \ref{sec:Summary} for more details). Here we include in the simulations the effects of additional energy sources that are required to unbind the envelope and estimate the merger rates in these scenarios. Moreover, we find different evolutionary routes that lead to NS/BH-core mergers and study the structure of the core during the merger event. 

Our study is organized as follows. We begin by describing the initial parameters and prescriptions we use in our population models (section \ref{sec:Model}). We then present the main evolutionary channels towards NS/BH-core mergers, the properties of binary systems that result in these mergers and their event rates (section \ref{sec:Results}). We summarize our findings and discuss their relevance to previous works (section \ref{sec:Summary}).

\section{Population synthesis model}
\label{sec:Model}

We use the population synthesis code Compact Object Mergers:  Population, Astrophysics and Statistics (\textsc{COMPAS}; \citealt{Stevensonetal2017}; \citealt{VignaGomezetal2018}; \citealt{Rileyetal2022}) to find the evolutionary pathways of binary systems that result in the merger of a NS/BH and the core of a giant star, explore their statistical properties and estimate the mergers' event rate. 

\textsc{COMPAS} generates populations of isolated stellar binary systems and evolves the stars in the binary using the analytical fits in \cite{Hurleyetal2000} based on the stellar models of \cite{Polsetal1998}. We sample the initial distribution of binary properties at the ZAMS (zero age main sequence) according to the ``Fiducial model'' of \cite{VignaGomezetal2018} as described below. Henceforth we refer to the initially heavier star as the primary star, and the initially lighter star as the secondary star. We draw the mass of the primary star from the Kroupa initial mass function (IMF) in the form $dN/dM_{\rm 1,ZAMS} \propto M_{\rm 1, ZAMS}^{-2.3}$ \citep{Kroupa2001} with masses in the range $ 5 \leq M_{\rm 1,ZAMS}/M_{\rm \odot} \leq 100$, and the mass of the secondary star from a flat distribution in the mass ratio with $0.1 \leq q_{\rm ZAMS} \equiv M_{\rm 2, ZAMS}/M_{\rm 1, ZAMS} \leq 1$ \citep{Sanaetal2012}. The initial separations follow the flat-in-the-log distribution between $ 0.1 \leq a_{\rm ZAMS}/\rm AU \leq 1000$ \citep{Sanaetal2012}. We take all of the orbits to be circular at ZAMS (i.e., $e_{\rm ZAMS}=0$), and all the stars in our sample have solar metallicity $Z=0.0142$  \citep{Asplundetal2009}.  

In general, massive stars end their lives in SN explosions. \textsc{COMPAS} differentiates between three SN scenarios: regular core collapse supernovae (CCSNe), electron capture supernovae (ECSNe) and ultra stripped supernovae (USSNe), according to the core and envelope masses prior to the explosion. We set the maximum allowed NS mass to be $M_{\rm NS,max} = 2 \, M_{\rm \odot}$, as according to \cite{ozeletal2010} this value reproduces the observational mass gap between NSs and BHs. For NS remnants, we take a bimodal natal-kick velocity distribution where CCSNe constitute the higher mode of $\sigma_{\rm high} = 265 \rm \; km s^{-1}$ and ECSNe together with USSNe contribute to the lower mode of $\sigma_{\rm low} = 30 \rm \; km s^{-1}$. We draw the BH natal kicks from the same bimodal distribution reduced according the fallback model of \cite{Fryeretal2012}.   

Throughout the evolution of a binary system the stars in the binary might interact through mass transfer. If the mass transfer is dynamically unstable a CEE phase begins (see \citealt{Rileyetal2022} for the implementation of mass transfer and its stability criteria in \textsc{COMPAS}). To determine the orbital separation between both stars in the binary system after CEE, \textsc{COMPAS} uses the energy formalism in which the energy difference between the orbital energies before and after the CEE phase is compared with the binding energy of the envelope (\citealt{vandenHeuvel1976}; \citealt{Webbink1984}; \citealt{LivioSoker1988}; \citealt{IbenLivio1993}; see \citealt{Ivanovaetal2013} for a review). In the case of a NS/BH that is swallowed by a giant star 
\begin{equation}
\begin{split}
 E_{\rm bind} = 
 \frac{\alpha_{\rm CE}GM_{\rm giant, pre-CE}M_{\rm NS/BH}}{2a_{\rm pre-CE}} \\ - \frac{ \alpha_{\rm CE}GM_{\rm giant, post-CE}M_{\rm NS/BH}}{2a_{\rm post-CE}}
\label{eq:EnergyFormalism}
\end{split}
\end{equation}
where $\alpha_{\rm CE}$ is the common envelope efficiency parameter, $M_{\rm NS/BH}$ is the mass of the compact object, $M_{\rm giant, pre-CE}$ and $M_{\rm giant, post-CE}$ are the masses of the giant star before and after the CEE phase respectively, and $a_{\rm pre-CE}$ and $a_{\rm post-CE}$ are the orbital separations before and after CEE respectively. The binding energy is calculated using the lambda formalism of \cite{deKool1990} as implemented by  \cite{XuLi2010,XuLi2010ERRATUM}. To find whether the NS/BH merges with the giant star's core, we compare the orbital separation of the systems after the CEE phase with the radius of the core, which is estimated using approximate analytical relations between the mass and radius of the core in different evolutionary phases from \cite{HallTout2014}. If the orbital separation is smaller than the radius of the core, we assume that a NS/BH-core merger has occurred. 

Traditionally, the maximal value of the common envelope efficiency parameter $\alpha_{\rm CE}$ is one, representing a case where the entire emitted orbital energy goes to unbind the envelope of the giant star and is precisely sufficient for this purpose. However, many hydrodynamical simulations do not find full envelope ejection, in contradiction to observations of post common envelope systems (e.g., \citealt{Passyetal2012}; \citealt{RickerTaam2012}; \citealt{Kuruwitaetal2016}; \citealt{Ohlmannetal2016}; \citealt{Iaconietal2017}; \citealt{GlanzPerets2021a}; \citealt{GlanzPerets2021b}), implying that additional energy sources are required to unbind the envelope. Several physical processes have been suggested as possible mechanism for the ejection of the common envelope, such as the core-companion system's interaction with a circumbinary disk in the final CEE stages (e.g., \citealt{KashiSoker2011}; \citealt{Kuruwitaetal2016}), jets launched from the compact object while it is inside the envelope of the secondary star (e.g., \citealt{Sabachetal2017}; \citealt{Soker2017}), the recombination energy of hydrogen and helium (e.g., \citealt{Ivanovaetal2015}; \citealt{Lauetal2022}), envelope inflation followed by long period pulsations (e.g., \citealt{Claytonetal2017}) and dust driven winds \citep{GlanzPerets2018}. In the energy formalism an additional energy source can be represented by $\alpha_{\rm CE}>1$ regardless of its nature.

We perform simulations of binaries with solar metllaicity for $0.1 \leq \alpha_{\rm CE} \leq 5$ to allow for both traditional CEE and scenarios with additional energy sources. For each value of $\alpha_{\rm CE}$, we evolve $10^{\rm 7}$ isolated binaries with the initial parameter distribution described above. In addition, we explore our results' dependence on metallicity by performing a grid of simulations with $Z= 10^{-4},10^{-3}, 2.5\times 10^{-2}$ over $\alpha_{\rm CE}=0.5,1,2$, and present our conclusions at the end of section \ref{subsec:NSBHcoreMergerEventRate}.

\section{Results}
\label{sec:Results}

\subsection{Constraining the values of $\alpha_{\rm CE}$}
\label{subsec:DoubleCompactObjectsMergerEventRate}

Due to the lack of ability to observe systems in the relatively short and low luminosity CEE phase (see \citealt{Ivanovaetal2013} for typical timescales of CEE) we cannot compare the results of our population synthesis simulations to direct observations. However, since roughly $80-90 \%$ progenitors of binary compact object mergers in our simulations go through the same evolutionary stages as progenitors of NS/BH-core mergers until the CEE of the compact objects with the giant star (Fig. \ref{fig:EvolutionRoutes}), we find their merger rates for different values of $\alpha_{\rm CE}$ and compare these rates with observations to determine the possible values of the efficiency parameter.

We crudely estimate the local transient rate density of different events in our COMPAS simulations using
\begin{equation}
R_{\rm event} = f_{\rm event} \frac{n_{\rm SFR} f_{\rm pop}}{\left <M\right >},
\label{eq:LocalRate}
\end{equation}
where $R_{\rm event}$ is the local transient rate density in $\rm Gpc^{-3} \yr^{-1}$, $f_{\rm event}=N_{\rm event} / N_{\rm total}$ is the ratio between the number of systems that result in this type of event and the total number of systems simulated in our COMPAS simulation, $n_{\rm SFR} \sim 10^{7} \, \rm M_{\rm \odot} Gpc^{-3} \yr^{-1}$ is the local star formation rate \citep{MadauDickinson2014}, $f_{\rm pop}$ is the fraction of the Kroupa IMF \citep{Kroupa2001} we simulate and $\left <M\right >$ is the average stellar mass. Integrating over the Kroupa IMF  between $M_{\rm min}=5 \, M_{\rm \odot}$ and $M_{\rm max}=100 \, M_{\rm \odot}$ and dividing by the total mass we find that $f_{\rm pop} \simeq 0.007$. The average stellar mass according to the Kroupa IMF is $\left <M\right > \simeq 0.39 \, M_{\rm \odot}$. Substituting all of the above into equation (\ref{eq:LocalRate}) gives 
\begin{equation}
R_{\rm event} = 1.79 \times 10^{5} f_{\rm event}  \rm \;  Gpc^{-3} \yr^{-1}.
\label{eq:LocalRateUnits}
\end{equation}

\begin{figure*}
\begin{center}
\vspace*{-0.5cm}
\hspace*{-0.3cm}
\includegraphics[width=0.7\textwidth]{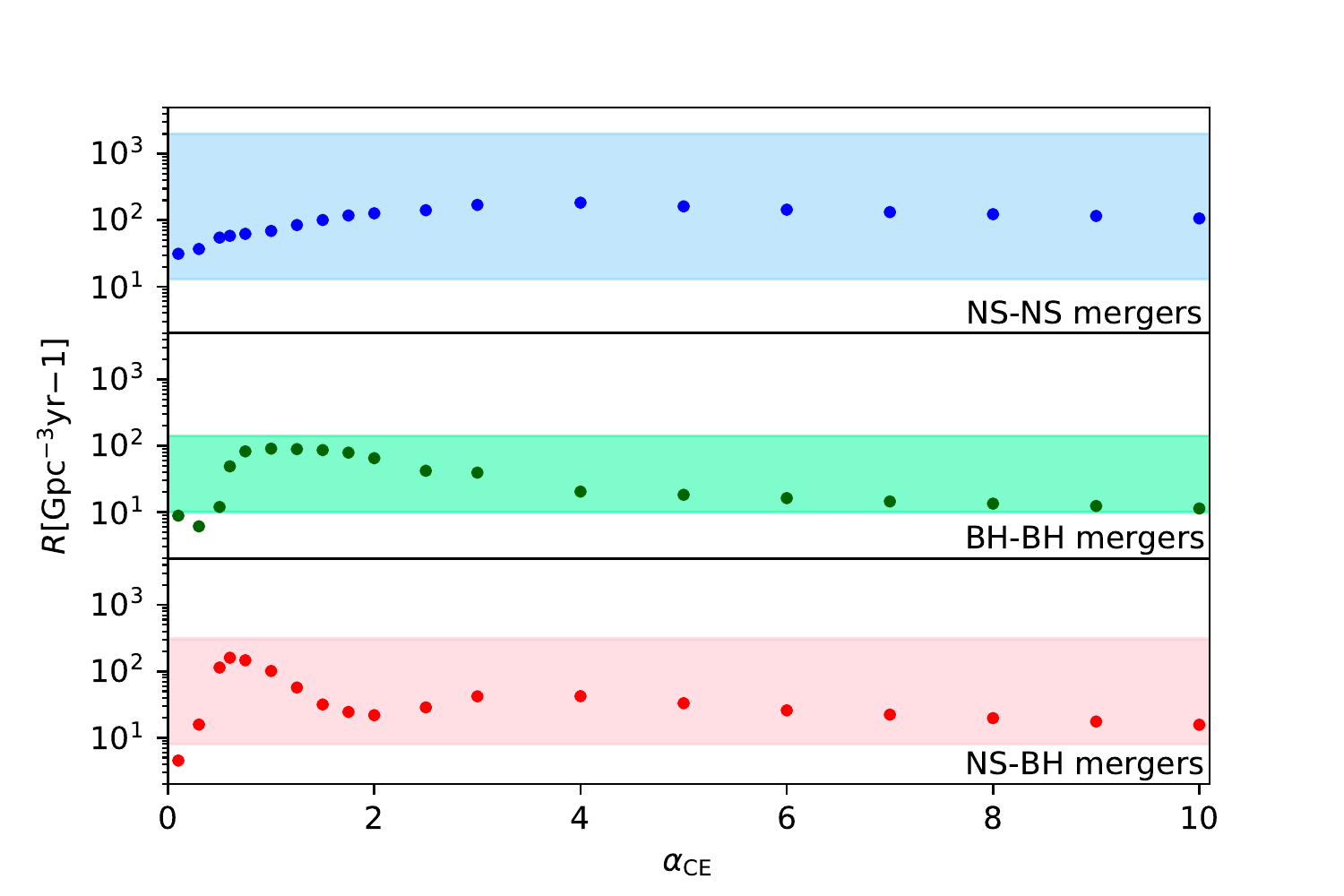}
\caption{Local rate density of NS-NS mergers (blue dots), BH-BH mergers (green dots) and NS-BH mergers (red dots) for different values of common envelope efficiencies. Statistical error bars are smaller than the marker size. The rectangular areas represent the error margin of mergers observation taken from \cite{Abbotetal2022}.  
}
\label{fig:CompactObjectsMergersRates}
\end{center}
\end{figure*}

In Fig. \ref{fig:CompactObjectsMergersRates} we present the local merger rate density of NS-NS binaries (upper panel; blue dots), BH-BH binaries (middle panel; green dots) and NS-BH binaries (lower panel; red dots) in our \textsc{COMPAS} simulations as computed by using equation (\ref{eq:LocalRateUnits}). The rectangular filled areas in each panel are the possible observational ranges of these rates as inferred from gravitational waves detected by the first three observing runs of the Advanced LIGO and the Advanced Virgo observatories (denoted as O1,O2 and O3) according to \cite{Abbotetal2022}. We conclude that the local merger rate densities are well within the observational error margin for $\alpha_{\rm CE} \gtrsim 0.5$. This is in accordance with previous studies which favour $\alpha_{\rm CE} \simeq 2$ compared to small $\alpha_{\rm CE}$ to explain the observed merger rates of double compact object binaries (e.g., \citealt{Garciaetal2021}; \citealt{Zevinetal2021}; \citealt{BroekgaardenBerger2021}).

\subsection{Evolutionary channels towards NS/BH-core mergers}
\label{subsec:Evolution}

\begin{figure*}
\begin{center}
\includegraphics[width=0.9\textwidth]{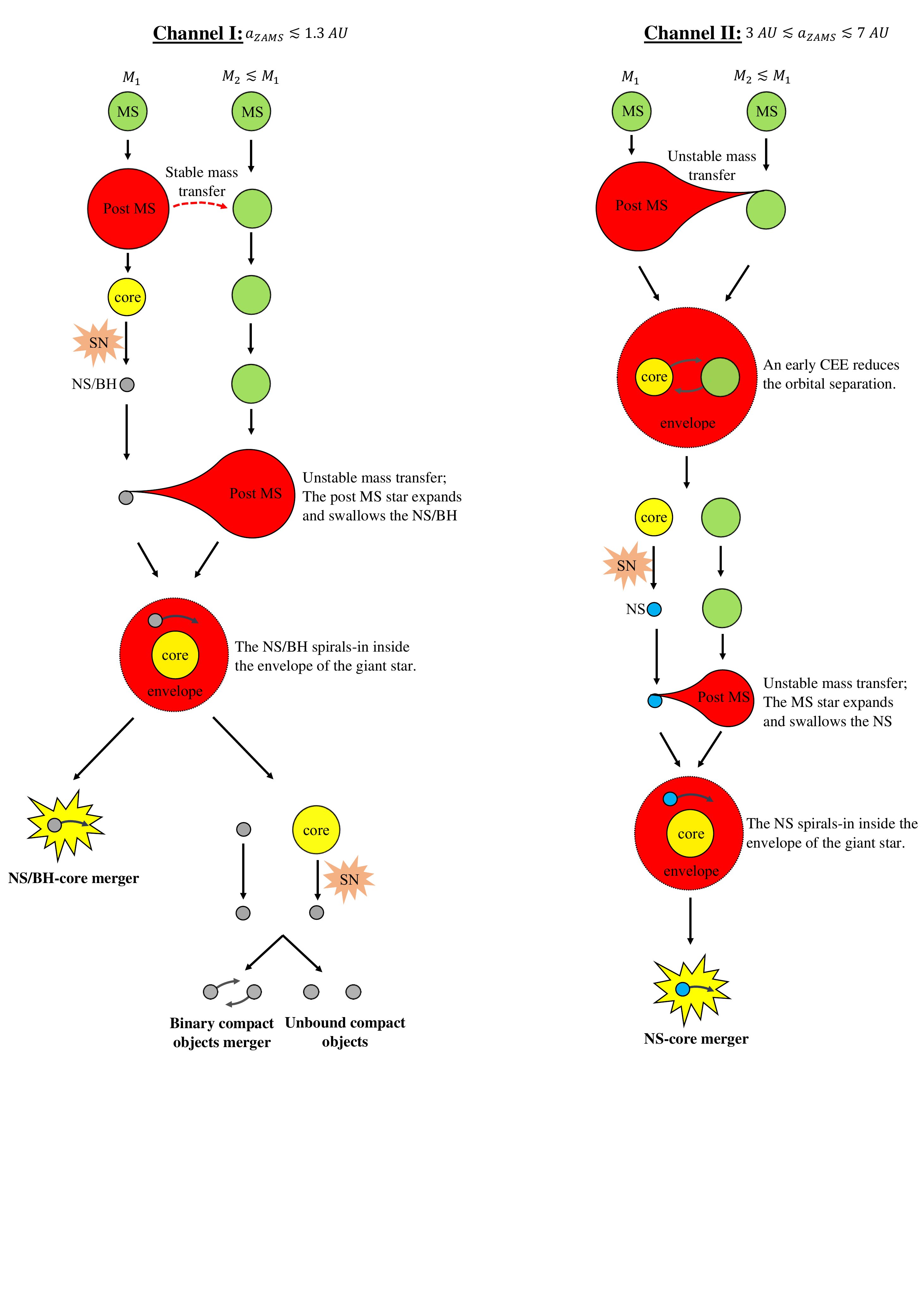}
\vspace*{-3.8cm}
\caption{A schematic illustration of the evolution of a massive binary system. Two massive MS stars evolve towards a giant star-NS/BH CEE. The NS/BH spirals-in inside the envelope of the giant star. The CEE can lead either to the merger of the NS/BH with the core of the giant star potentially powering a bright transient event, or to the formation of double compact objects binary that might merge in the future by emission of gravitational waves. Left panel: main evolution channel, where dynamically stable mass transfer from the post MS primary star to its MS secondary companion takes place. Right Panel: Secondary evolution channel in which a dynamically unstable mass transfer leads to the formation of a common envelope that surrounds the secondary MS star and the core of its post MS primary, reducing their orbital separation, and allowing only for NS-core mergers. 
}
\label{fig:EvolutionRoutes}
\end{center}
\end{figure*}

The two main evolutionary channels of binary systems that result in the merger of a NS/BH with the core of a giant star are presented in Fig. \ref{fig:EvolutionRoutes}. Both scenarios begin with two massive stars on the MS. The initially heavier star $M_{\rm 1}$, to which we will refer as the primary star, evolves faster and becomes a giant while the lighter secondary star $M_{\rm 2}$ keeps burning hydrogen at its core. 

In the main channel, denoted as \textit{channel I} (left panel of Fig. \ref{fig:EvolutionRoutes}), a stable mass transfer episode from the giant primary to its MS secondary occurs at this stage. The giant star eventually loses its envelope through mass transfer and the remaining naked core keeps evolving until it explodes in an stripped-envelope SN event in the same manner as Wolf-Rayet stars\footnote{We note that the stripped core in our scenario is not necessarily a Wolf-Rayet star according to the observational definition \citep{Shenaretal2020}.} (e.g., \citealt{WheelerLevreault1985}; \citealt{Woosleyetal1995}; \citealt{EldridgeTout2004}; \citealt{McClellandEldridge2016}), leaving a NS/BH remnant, depending on the mass of the stripped core. The natal kick might drive the compact object away from the SN location, but a low enough kick allows for a bound NS/BH-MS binary system (e.g., \citealt{Kochaneketal2019}). The secondary star continues to evolve and when the hydrogen in its core is depleted it enters the Hertzsprung Gap (HG) phase where it expands until the ignition of helium in its core. From the HG phase and onward, mass transfer occurs in the opposite direction, i.e. from the secondary star to the primary. The expanding secondary becomes a giant and it can engulf the NS/BH initiating a CEE phase where the compact object spirals-in inside the envelope of the giant star. The compact object might be swallowed by the giant star later in the evolution as well, e.g, during core helium burning, as we show in Fig. \ref{fig:PercentHeCO}. Tidal forces can bring the NS/BH into the envelope even if the binary separation is several times larger than the radius of the giant star (e.g., \citealt{Soker1996}). Currently the \textsc{COMPAS} code does not include tidal evolution, implying that it underestimates the number of systems that go through NS/BH-giant star CEE, and therefore the rates of NS/BH core mergers and their resultant transients.

The giant star-NS/BH CEE has two possible outcomes. If the envelope is not ejected before the NS/BH enters the core then the NS/BH merges with the core of the giant star as shown on the left in the left panel of Fig. \ref{fig:EvolutionRoutes}. However, if the envelope is entirely unbound before the NS/BH reaches the core, then the core will keep evolving and eventually ends its life in a stripped-envelope SN explosion resulting in another NS/BH (right side in the left panel of Fig. \ref{fig:EvolutionRoutes}). If the binary remains bound after the second SN explosion, and the two compact objects are close enough, they might merge on a timescale shorter than the Hubble time emitting potentially detectable gravitational waves (the majority of systems shown in Fig. \ref{fig:CompactObjectsMergersRates}). \cite{VignaGomezetal2018} find that this formation channel is responsible for roughly $70 \%$ of the Galactic binary NSs. 

Less massive and wider binaries evolve through a secondary channel (\textit{channel II}; right panel of Fig. \ref{fig:EvolutionRoutes}), where dynamically unstable mass transfer between the post MS primary star and its MS secondary leads to an early CEE phase that brings the core of the giant and the MS star closer together. At the end of CEE the envelope is ejected and the evolution continues as in \textit{channel I}. We note that this formation channel can only involve a NS compact object due to the relatively low masses of the SN progenitor. A massive BH progenitor always leads to stable mass transfer from the post-MS primary to its MS secondary (second stage of channel I in Fig. \ref{fig:EvolutionRoutes}), implying that all BH-core mergers evolve through \textit{channel I}. This evolution channel always ends with the merger of the NS with the core of the giant star due to the small orbital separation at the end of first CEE (see equation \ref{eq:EnergyFormalism}). In both formation channels presented in Fig. \ref{fig:EvolutionRoutes}) several short dynamically stable mass transfer episodes might occur throughout the evolution besides the mass transfer events mentioned above.

\begin{figure}
\begin{center}
\vspace*{-0.7cm}
\hspace*{-0.3cm}
\includegraphics[width=0.52\textwidth]{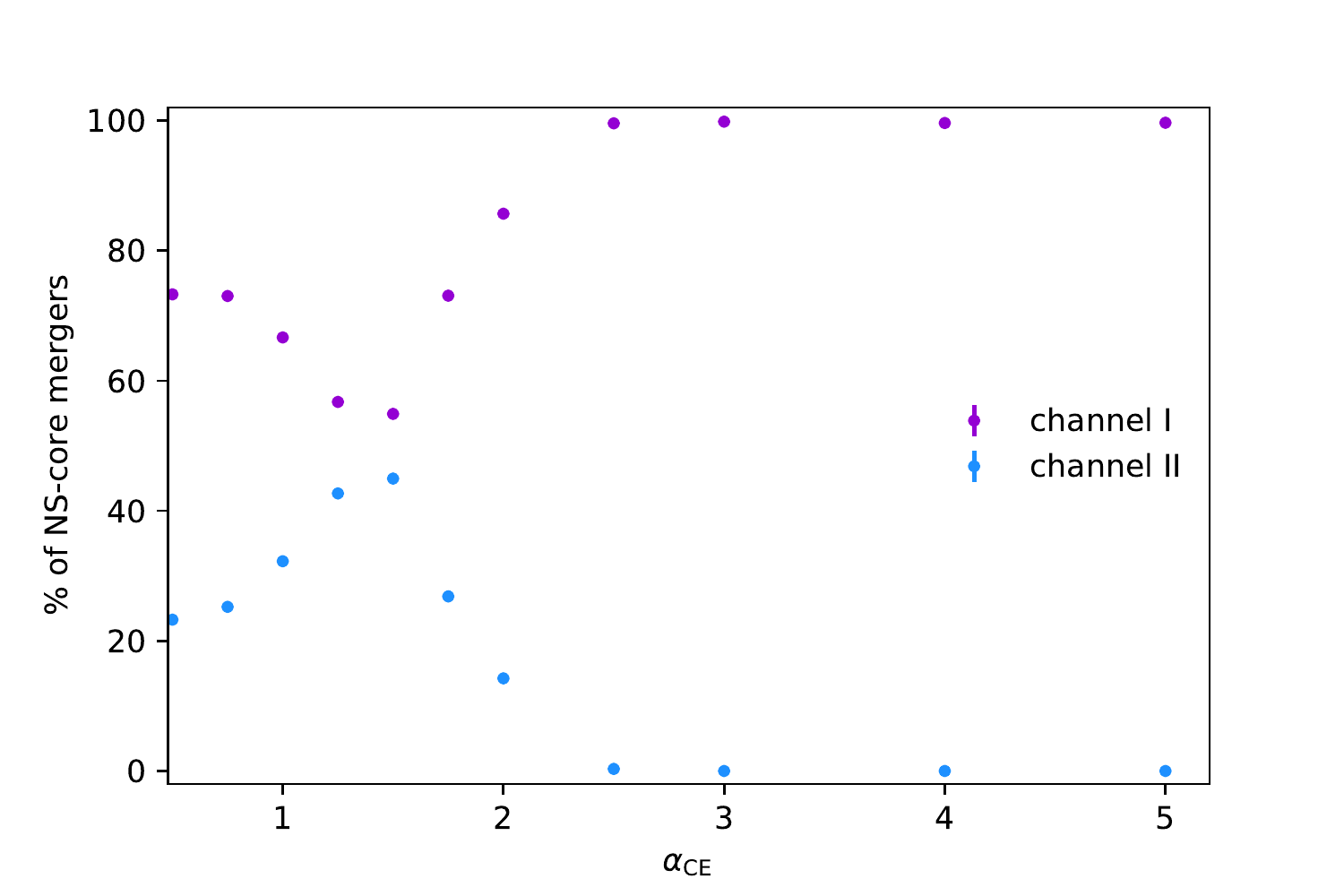}
\caption{Percentage of NS-core mergers that evolved through \textit{channel I} (purple dots) and \textit{channel II} (blue dots) from all NS-core mergers for values of $\alpha_{\rm CE}$ that coincide with observations of double compact object mergers (Fig. \ref{fig:CompactObjectsMergersRates}). Statistical error bars are smaller than the marker size.
}
\label{fig:PercentChannels}
\end{center}
\end{figure}

Fig. \ref{fig:PercentChannels} presents the percentage of systems that evolve through \textit{channel I} (purple dots; left panel of Fig. \ref{fig:EvolutionRoutes}) and through \textit{channel II} (blue dots; right panel of Fig. \ref{fig:EvolutionRoutes}) for different values of $\alpha_{\rm CE}$ in our simulations. A remaining small percentage of the systems evolved through various other channels, including binary evolution where there is no mass transfer between the stars in the binary system when the primary star is a giant. We note a non-monotonic behaviour with $\alpha_{\rm CE}$ as a result of two separate CEE phases that are affected by this parameter. A larger value of $\alpha_{\rm CE}$ implies a larger orbital separation in the end of CEE. In \textit{channel II} this means that on the one hand more systems survive the first CEE phase (between the giant primary and the MS secondary) and therefore can lead to NS-core mergers, but on the other hand in the second CEE (NS primary-giant secondary) more NSs do not manage to enter the core of the giant star.

\subsection{Properties of binary systems that result in NS/BH-core mergers}
\label{subsec:Properties}
\begin{figure*}[h]
\begin{center}
\vspace*{-1.3cm}
\includegraphics[width=0.80\textwidth]{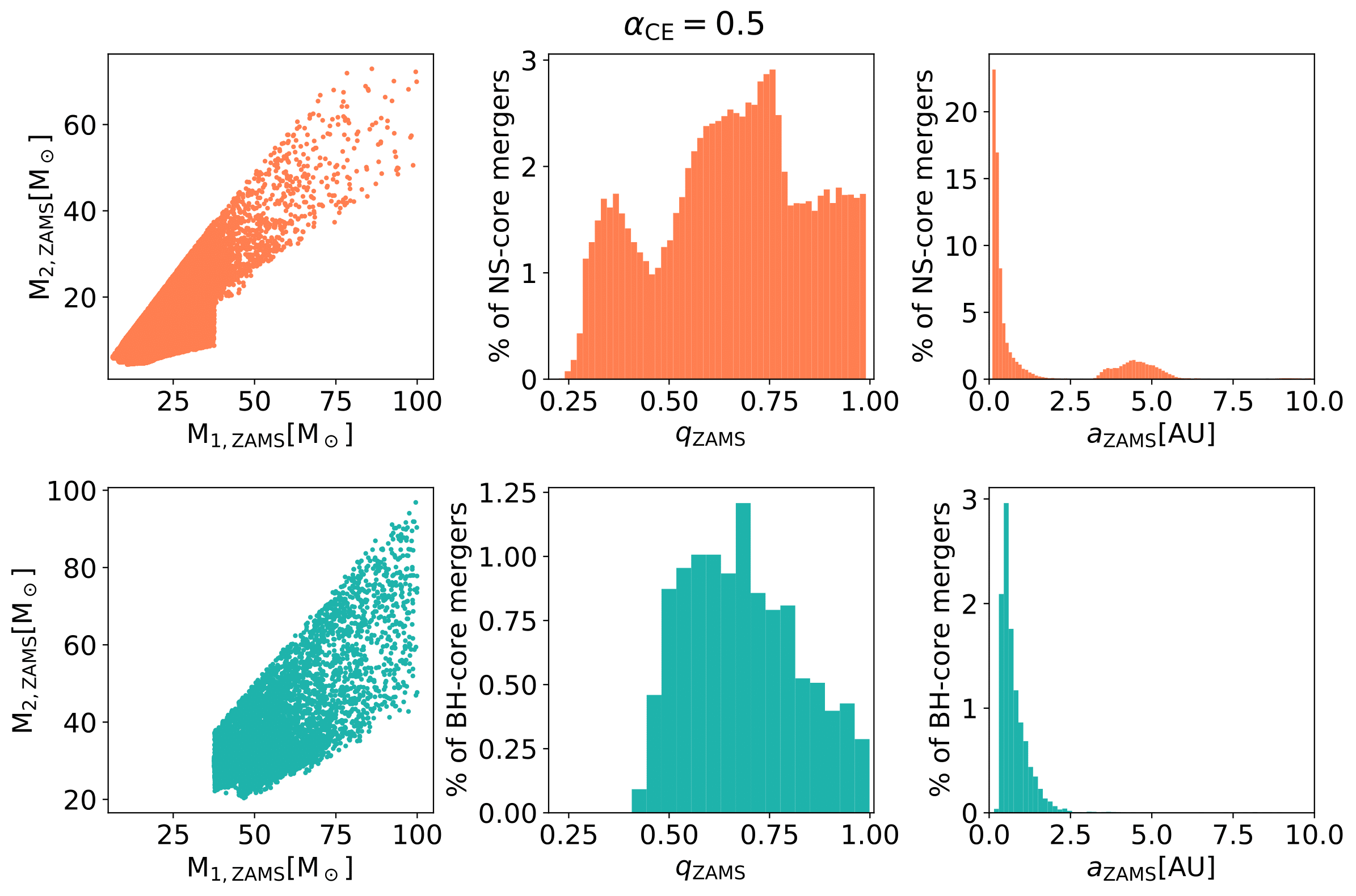}
\caption{Distributions of binary properties at the ZAMS of systems that result in NS-core mergers (upper panels; orange) or in BH-core mergers (lower panels; turquoise) for a common envelope efficiency parameter $\alpha_{\rm CE}=0.5$. Left panels: initial mass of the NS/BH (orange dots/turquoise dots) progenitors (primary star, $M_{\rm 1,ZAMS}$) vs the initial mass of the star that swallows the NS/BH (secondary star, $M_{\rm 2,ZAMS}$). Middle panels: initial mass ratio $q_{\rm ZAMS} \equiv M_{\rm 2, ZAMS}/M_{\rm 1, ZAMS}$ that leads to NS/BH-core mergers. The orange bins represent the percentage of binary systems that begin with a certain mass ratio and result in a NS-core merger from all systems where the compact object merges with the core of the giant star during the CEE phase. The width of each bin is $0.015$. The turquoise bins, whose width is $0.037$ depict this percentage for BH-core mergers.  Right panels: Similar to the middle panels for the binary initial orbital separations. In these distributions the widths of the orange and turquoise bins are $0.1 \rm \,  AU$, and $0.15 \rm \,  AU$, respectively.  
}
\label{fig:alpha0.5}
\end{center}
\end{figure*}

\begin{figure*}
\begin{center}
\vspace*{-0.5cm}
\includegraphics[width=0.80\textwidth]{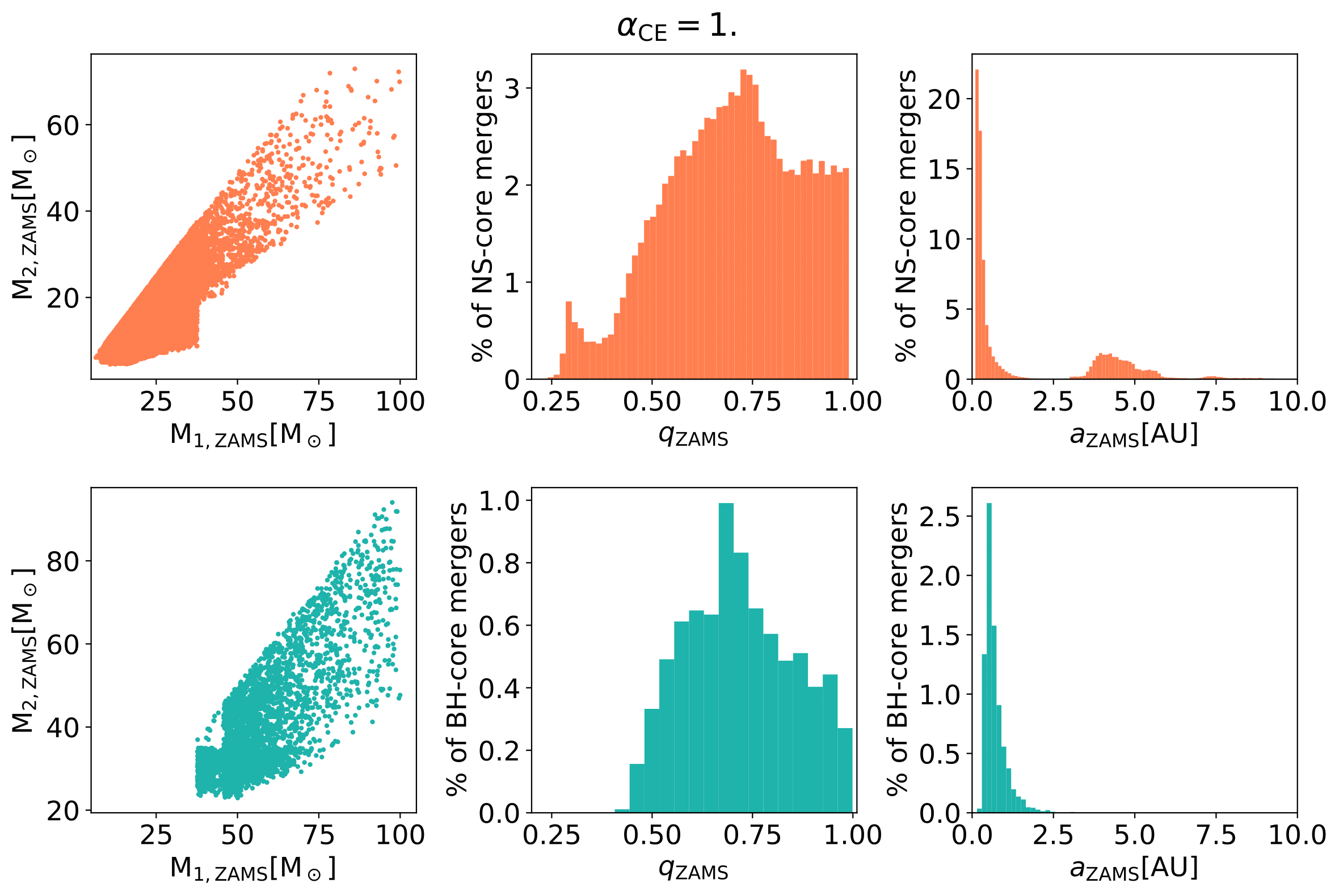}
\caption{Similar to Fig. \ref{fig:alpha0.5} for a common envelope efficiency parameter $\alpha_{\rm CE}=1$.
}
\label{fig:alpha1}
\end{center}
\end{figure*}

\begin{figure*}
\begin{center}
\vspace*{-0.5cm}
\includegraphics[width=0.80\textwidth]{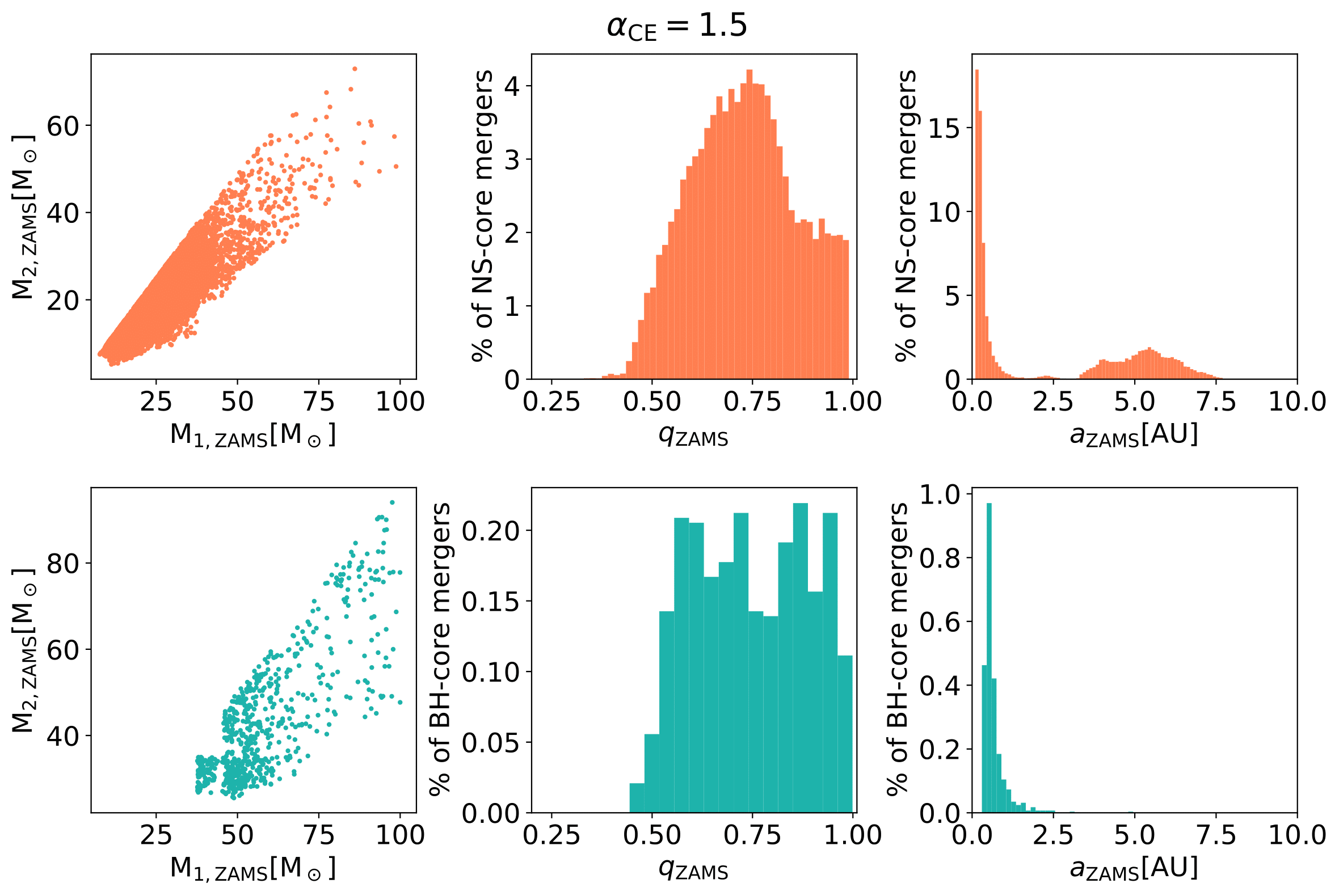}
\caption{Similar to Fig. \ref{fig:alpha0.5} for a common envelope efficiency parameter $\alpha_{\rm CE}=1.5$.
}
\label{fig:alpha1.5}
\end{center}
\end{figure*}

\begin{figure*}
\begin{center}
\vspace*{-1.5cm}
\includegraphics[width=0.80\textwidth]{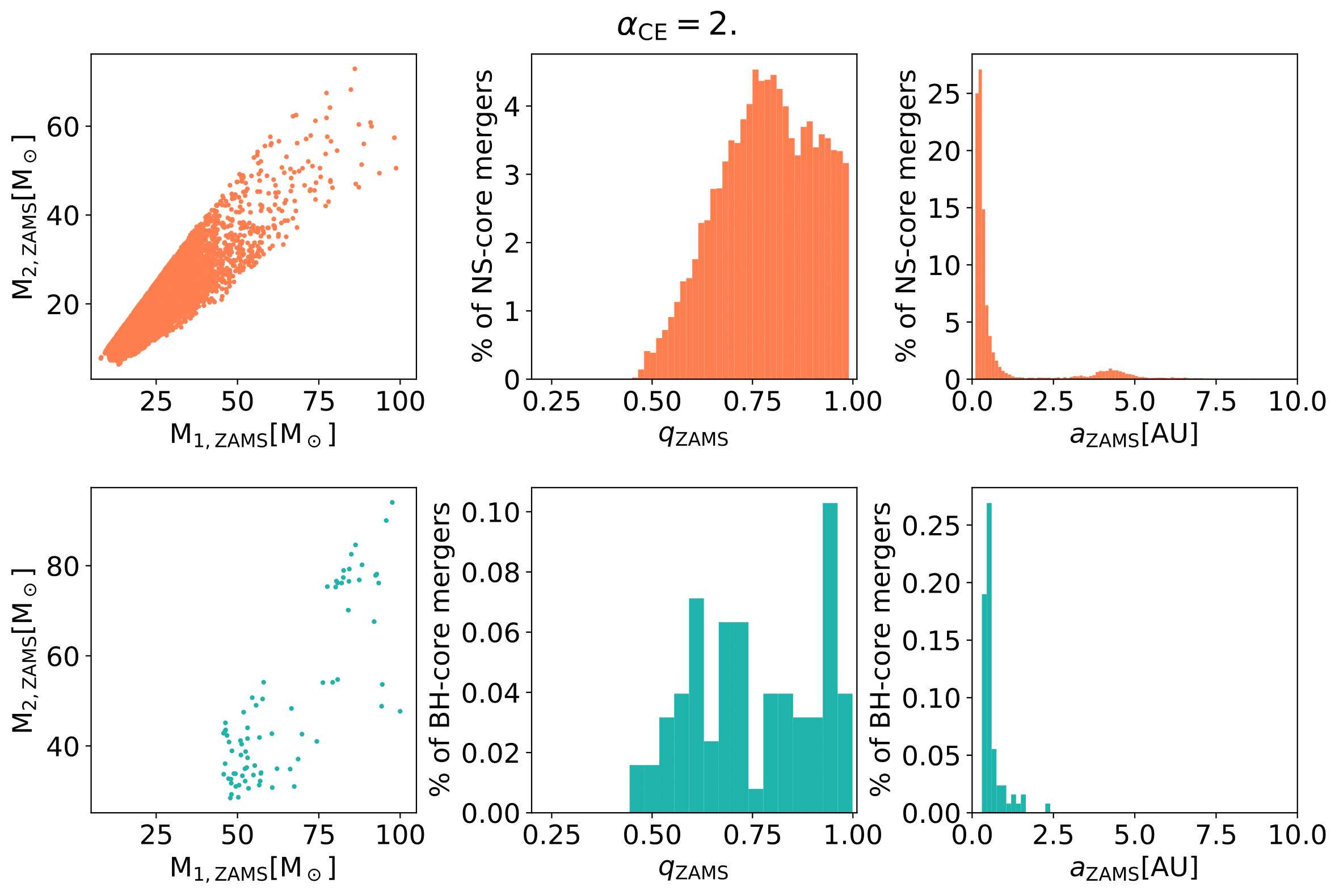}
\caption{Similar to Fig. \ref{fig:alpha0.5} for a common envelope efficiency parameter $\alpha_{\rm CE}=2$.
}
\label{fig:alpha2}
\end{center}
\end{figure*}

As mentioned in section \ref{sec:Model}, we use \textsc{COMPAS} to explore the properties of NS/BH-core merger progenitors. In Figs. \ref{fig:alpha0.5}-\ref{fig:alpha2} we show different distributions of these progenitors at the ZAMS for representative values of $\alpha_{\rm CE}$.
Several interesting trends emerge from these figures. While NS-core mergers occur for $M_{\rm 1, ZAMS} \gtrsim 6.5 \, M_{\rm \odot}$ up to the heaviest stars in our mass distribution, BH-core mergers tend to involve much heavier stars in the initial binary system, with $M_{\rm 1, ZAMS} \gtrsim 37.7 \, M_{\rm \odot}$ (left panels of Figs. \ref{fig:alpha0.5}-\ref{fig:alpha2}). This can be explained by the heavier masses required to produce a BH in a SN event under the assumptions of \textsc{COMPAS}. Binary systems with higher mass ratios at the ZAMS are more likely to result in NS/BH-core mergers. For both mergers of NSs and BHs with cores of giant stars most of the progenitor binaries have initial mass ratios of $q_{\rm ZAMS}  \equiv M_{\rm 2, ZAMS}/M_{\rm 1, ZAMS} \gtrsim 0.5$ (middle panels of Figs. \ref{fig:alpha0.5}-\ref{fig:alpha2}), while NS-core mergers can also originate from $0.25 \lesssim q_{\rm ZAMS} \lesssim 0.5$. The percentage of systems of the latter mass ratio strongly decreases with $\alpha_{\rm CE}$. 

We note a bimodal distribution in the initial orbital separation of systems that result in a core-NS merger (right upper panels of Figs. \ref{fig:alpha0.5}-\ref{fig:alpha2}) where each peak corresponds to one of the evolution channels presented in Fig. \ref{fig:EvolutionRoutes}. We find that the higher/lower peak in the initial separation coincide with the percentage of systems that evolve through channel I/II (Fig. \ref{fig:PercentChannels}).

The larger population is in small separations and has a peak at $a_{\rm ZAMS} \simeq \rm 0.65 \, AU \simeq 130 \, R_{\rm \odot}$ for most values of $\alpha_{\rm CE}$ we simulated. It corresponds to \textit{channel I}, which is characterized by stable mass transfer from the post MS primary to the MS secondary. One might expect that small initial separation would typically lead to unstable mass transfer and CEE, and this is indeed a common outcome in our population models. However, in most such systems the secondary star merges with the core of the giant primary while still on the MS and hence cannot lead to a NS/BH-core merger, i.e. the prevalence of stable mass transfer we find at small separations is a selection effect. The stars in binary systems whose initial orbital separation is sufficiently small remain close enough so that the resulting NS can be swallowed by the secondary at its giant phase and eventually merge with its core, while for systems that begin further apart the NS and the giant star do not engage in CEE. BH-core mergers exhibit only the first peak in the initial orbital separation, i.e., evolve only through \textit{channel I}\footnote{We made sure the single peak is not due to a small sample of BH-core mergers by performing simulations of $10^7$ binaries drawing the mass of the primary from Kroupa IMF with masses in the range $ 35 \leq M_{\rm 1}/M_{\rm \odot} \leq 100$. This is equivalent to running $1.6 \times 10^8$ systems in the range $ 5 \leq M_{\rm 1}/M_{\rm \odot} \leq 100$. }.  

The second population of binaries that end their lives in a merger of a NS with the core of a giant star during CEE is smaller and has a peak around $a_{\rm ZAMS} \simeq \rm 4.65 \, AU \simeq 1000 \, R_{\rm \odot}$. This population corresponds to \textit{channel II}. Primaries with masses in the range $6.5 \, M_{\rm \odot} \lesssim M_{\rm 1, ZAMS} \lesssim 20 \, M_{\rm \odot}$ evolve through this channel. In this case a common envelope is formed around the core of the post MS primary star and the MS secondary, and dynamical friction between the core-MS system and the gas of the envelope reduces the orbital separation significantly \footnote{As there is still no analytical formulation for dynamical friction, the reduction in orbital separation during CEE in \textsc{COMPAS} is handled through the energy formalism.}. At some point the common envelope is ejected and a bound, more compact system of a MS star and the naked core remains behind, and keeps evolving towards a NS-core merger as in \textit{channel I}. Even though stable mass transfer can occur between the post MS primary and the MS secondary in this regime of initial orbital separations, the stars in such binaries are not close enough to lead to a CEE between the NS and the giant at a later stage, and therefore cannot lead to a NS-core merger event. For even larger initial orbital separations stars in the binary systems would mainly evolve separately.

\subsection{Mergers' event rate}
\label{subsec:NSBHcoreMergerEventRate}

\begin{figure*}
\begin{center}
\includegraphics[width=0.7\textwidth]{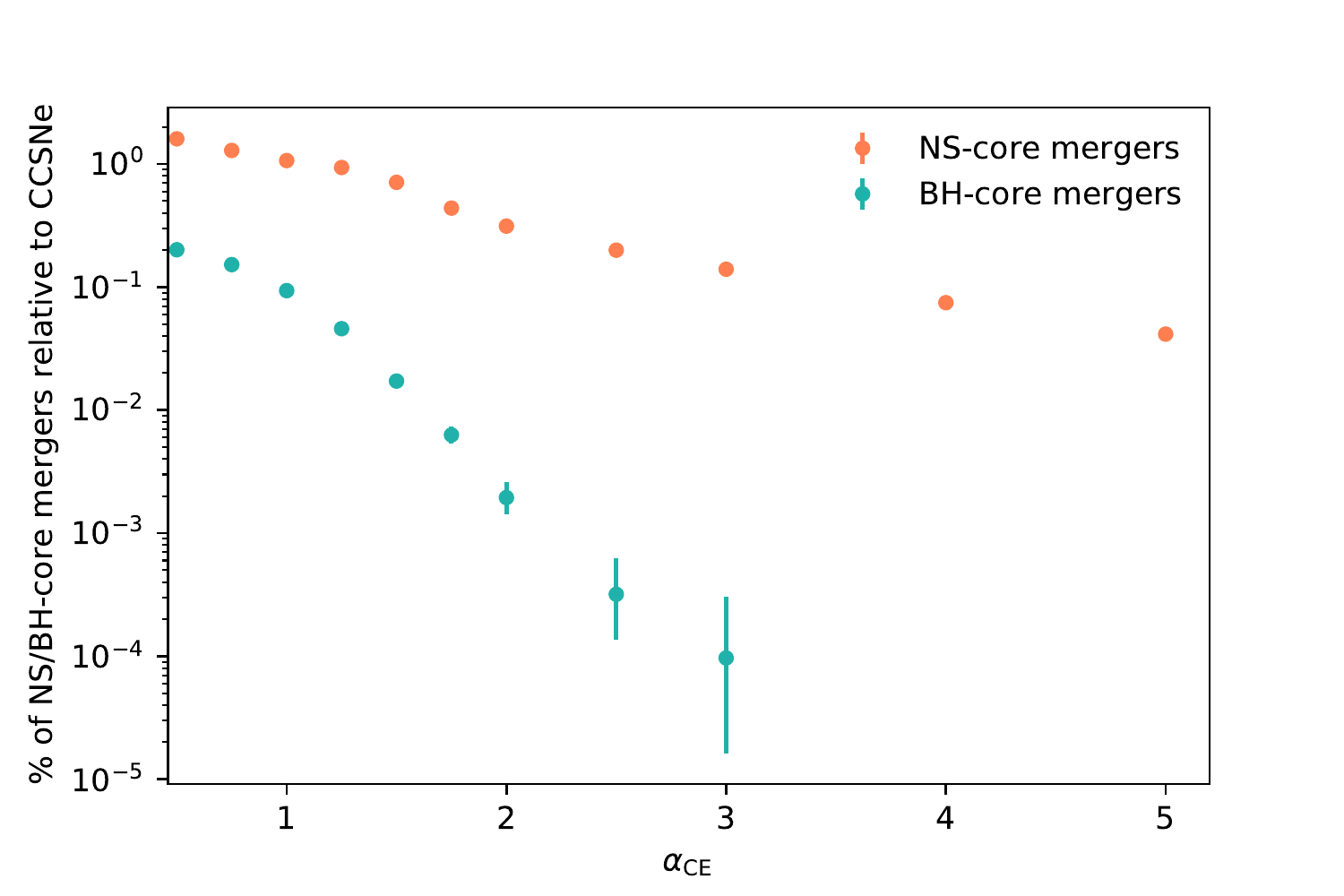}
\caption{Percentage of NS-core mergers (orange dots) and BH-core mergers (turquoise dots) of all CCSNe for the values of common envelope efficiencies that coincide with observations of binary compact object mergers (Fig. \ref{fig:CompactObjectsMergersRates}). The error bars represent statistical uncertainty. 
}
\vspace*{-0.5cm}
\label{fig:CompactObjectCoreMergersRates}
\end{center}
\end{figure*}

We can estimate the merger rates of NSs/BHs with the cores of the giant secondaries using the results of our population synthesis models. We present the percentage of these rates relative to CCSNe events in Fig. \ref{fig:CompactObjectCoreMergersRates} for the values of $\alpha_{\rm CE}$ we find in section \ref{subsec:DoubleCompactObjectsMergerEventRate}. The orange and turquoise dots are the percentages of NS-core and BH-core mergers, respectively, from CCSNe in our population model. 

We note that for both NS and BH mergers with the giant's cores the event rate decreases with $\alpha_{\rm CE}$, a trend we can also see by comparing the left panels of Figs. \ref{fig:alpha0.5}-\ref{fig:alpha2}. In the energy formalism (see equation \ref{eq:EnergyFormalism}), higher values of $\alpha_{\rm CE}$ imply that more energy is transmitted from the orbit (or from an additional energy source) to the envelope and it is ejected earlier in the evolution. The underlying physical concept is that if a larger amount of energy is deposited inside the envelope it expands to larger radii before the outer layers are ejected, becoming less dense and reducing the dynamical friction between the spiraling-in primary and the gas of the secondary's envelope. This results in larger orbital separations at the end of CEE, implying that more NSs/BHs remain outside the core rather than merge with it. BH-core mergers reduce to nearly nothing for $\alpha_{\rm CE} \gtrsim 4$ while the event rate of NS-core mergers drops about an order of magnitude between the lowest and highest values of   $\alpha_{\rm CE}$ we simulate.

\begin{figure*}
\begin{center}
\vspace*{-0.7cm}
\hspace*{-0.5cm}
\includegraphics[width=0.7\textwidth]{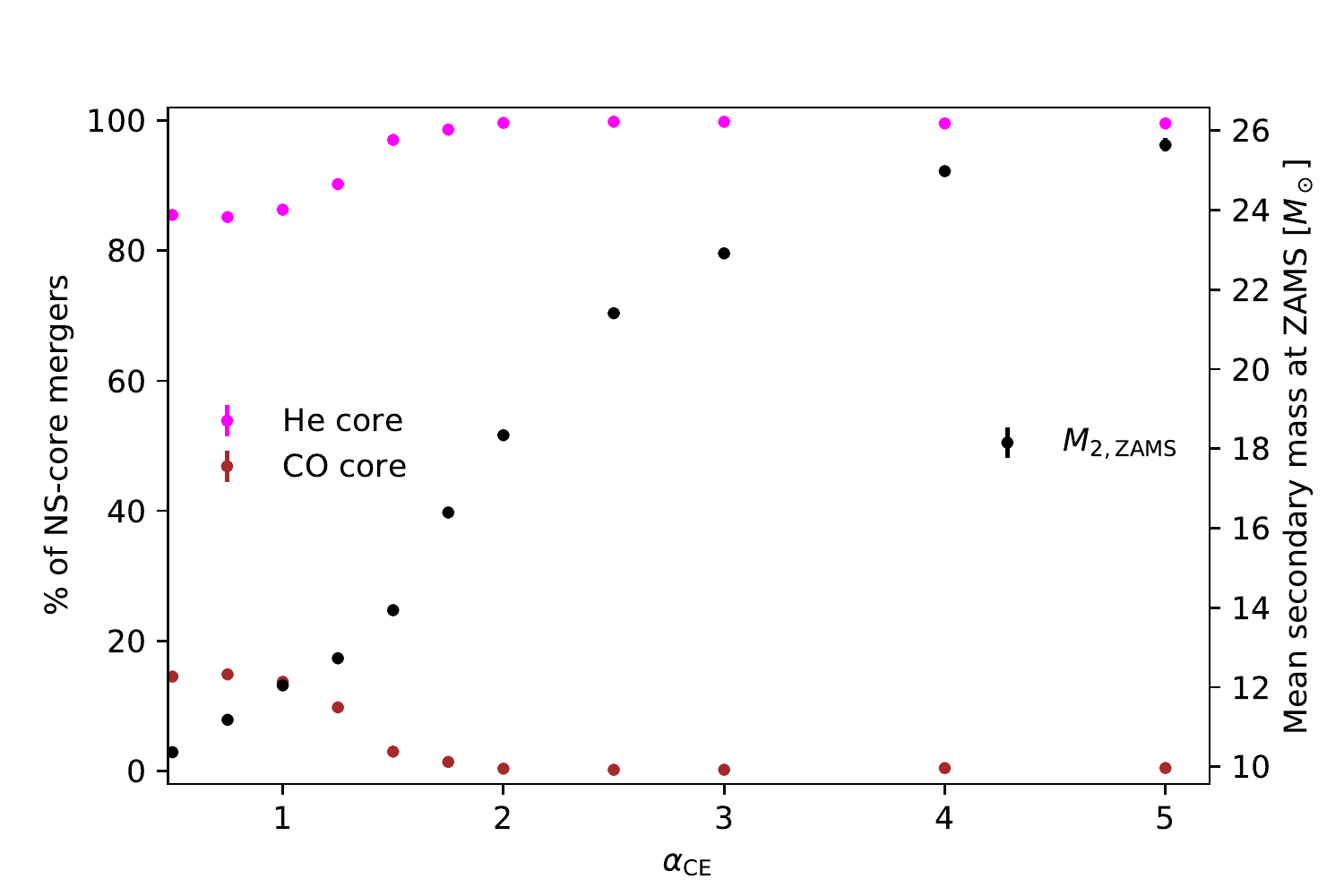}
\caption{Percentage of NS-core mergers where the core of the giant secondary star is composed of helium (magenta dots) or Carbon-Oxygen (brown dots) from all NS-core mergers for different values of the efficiency parameter $\alpha_{\rm CE}$. The black dots are the average mass of the secondary at the ZAMS for each value of $\alpha_{\rm CE}$. Most statistical error bars are smaller than the marker size. 
}
\label{fig:PercentHeCO}
\end{center}
\end{figure*}

In Fig. \ref{fig:PercentHeCO} we present how the percentage of systems where a NS merges with the helium (magenta dots) or carbon-oxygen (brown dots) core of a giant star from all NS-core mergers varies with the efficiency parameter $\alpha_{\rm CE}$. For larger values of $\alpha_{\rm CE}$ the number of mergers where the core is composed of helium is overall larger, i.e., the mergers occur at earlier times during the giant-NS/BH CEE. As larger amounts of energy go to unbind the envelope it is ejected at an earlier stage, and the orbital separation at the end of the CEE phase is larger (equation \ref{eq:EnergyFormalism}). This implies that mergers between the giant's core and the NS for larger values of $\alpha_{\rm CE}$ occur for more massive secondaries with higher envelope binding energies (as we indeed see in the trend of the black dots in \ref{fig:PercentHeCO}). Such giants are larger to begin with and thus can swallow the NS as they expand at earlier stages, before core helium burning. We find that binary systems whose outcome is BH-core mergers require $\left <M_{\rm 2, ZAMS} \right> \gtrsim 37.7 \, M_{\rm \odot}$ (section \ref{subsec:Properties}). Therefore, all BH-core mergers occur when the core of the giant secondary is composed mainly of helium regardless of the $\alpha_{\rm CE}$ value. 

We examine the effects of mass accretion by the NS/BH while inside the envelope of the giant star and find no statistically significant differences in any of the quantities we have discussed due to relatively low mass accretion rates (appendix). Moreover, to explore the effect of metallicity on our results we performed a grid of simulations with $Z= 10^{-4},10^{-3}, 2.5\times 10^{-2}$ over $\alpha_{\rm CE}=0.5,1,2$. We find that while the rate of NS-core mergers does not vary much with metallicity, there are less BH-core mergers in higher values of $Z$, up to an order of magnitude difference between the highest and lowest metallicity we explored. We attribute these findings to the increase of stellar mass loss with metallicity (e.g., \citealt{NieuwenhuijzendeJager1990}; \citealt{Kudritzkietal1989}) that leads to less massive stellar remnants after the SN explosion in higher values of $Z$, i.e., less BHs are formed in higher metallicites. This leads to less BH-core mergers, but does not affect much the NS-core merger rates due to the much larger number of NSs formed in the simulations relative to BHs.

\section{Summary and discussion}
\label{sec:Summary}

In this work we performed population synthesis of massive binaries in search of NSs and BHs mergers with cores of giant stars. We used \textsc{COMPAS} to generate populations of massive binary systems (section \ref{sec:Model}) and followed the evolution of binaries that result in these mergers (section \ref{subsec:Evolution}). We used the energy formalism to determine whether a merger occurred during the CEE of the giant secondary star with the NS/BH. We simulated cases of traditional CEE in which the orbital energy released due to the decay in the orbit is the main cause for envelope ejection ($\alpha_{\rm CE} \le 1$), and other cases where we assume an additional energy source ($\alpha_{\rm CE} > 1$). We constrained the possible values of $\alpha_{\rm CE}$ by comparing observations of binary compact object mergers to our simulated rates (section \ref{subsec:DoubleCompactObjectsMergerEventRate}).  

We found one main evolution route of NS-core and BH-core merger events (\textit{channel I}; left panel of Fig. \ref{fig:EvolutionRoutes}; same evolution channel as in \citealt{Schroderetal2020}) and an additional secondary evolution pathway for NS-core mergers (\textit{channel II}; right panel of Fig. \ref{fig:EvolutionRoutes}). The right panels of Figs. \ref{fig:alpha0.5}-\ref{fig:alpha2} exhibit a bimodal distribution in the initial orbital separation of NS-core merger progenitors whose modes are consistent with the two different evolution routes. For relatively close binaries (with initial orbital separation $a_{\rm ZAMS} \lesssim 1.3\, AU$) only cases where dynamically stable mass transfer occurs when the primary star is in its post MS phase can lead to NS (and also BH)-core mergers (\textit{channel I}). However, at larger initial orbital separations ($3 \, AU \lesssim a_{\rm ZAMS} \lesssim 7 \, AU$) a CEE is required to reduce the separation of the core-MS binary and bring them close enough to allow for a later CEE between the NS and the giant in which the NS might merge with its core (\textit{channel II}).  

We compute the mergers' event rates and find their percentage relative to CCSNe for different values of the CEE efficiency parameter we simulated (section \ref{subsec:NSBHcoreMergerEventRate}). We found that for $\alpha_{\rm CE}=1$, which is the commonly used value for CEEs that involve massive giants, there is about 1 NS-core merger per 100 CCSN events and about 1 BH-core merger per 1000 CCSN events. For nearly the same parameters space \cite{Schroderetal2020} find about 3 merger events per 1000 CCSNe for both NS and BH-core mergers. From a comparison between the percentage of systems that go through each stage of the evolution towards NS/BH-core merger, we attribute the difference in the results mainly to changes in the way different versions of \textsc{COMPAS} handle the entrance to CEE. 

As we can see in Fig. \ref{fig:CompactObjectCoreMergersRates}, the number of mergers decreases with the CEE efficiency parameter $\alpha_{\rm CE}$ as expected from the energy formalism and its underlying physics. For $\alpha_{\rm CE} \gtrsim 4$ the event rate of BH-core mergers reduces to nearly nothing within the accuracy of our simulations.  We estimate the amount of accreted mass by the NS and by the BH while inside the envelope of the giant star in the CEJSN scenario (appendix) and find it does not affect the mergers' event rate. We explored the effect of metallicity on our results and found that while changes in the systems' metallicities do not affect much the merger rates of NSs with cores of giant stars, there is an order of magnitude difference between the rates of BH-core mergers in the highest and in the lowest metallicities in our simulations.

The energy formalism for CEE has several shortcomings. The large uncertainty in the energy and mass transfer during the CEE phase might greatly affect the rates of transients whose progenitor binaries go through CEE (e.g., \citealt{Olejaketal2021}). Moreover, many hydrodynamical simulations find that the orbital energy cannot be the sole energy source that contributes to envelope ejection in massive binaries (see references in section \ref{sec:Model}). Using values of $\alpha_{\rm CE}>1$ to represent additional energy sources, as we did in the present study, disregards that these energy sources do not depend on the orbital separation in the same way as the orbital energy. However, keeping in mind this formalism is a phenomenological treatment, $\alpha_{\rm CE}>1$ yields that the spiraling-in star can end at a larger radius, and allows to obtain reasonable results in population synthesis of massive stars. A possible refinement of our analysis would take into account that $\alpha_{\rm CE}$ is in general system-dependant. For instance, if jets are the additional energy source that unbinds the envelope $\alpha_{\rm CE}$ is naturally larger for a NS accretor than a MS star, and even larger for a BH. Another example is the study of \citealt{DeMarcoetal2011} which finds that $\alpha_{\rm CE}$ is inversely correlated with the binary mass ratio. We note that other prescriptions of CEE applicable to population synthesis, such as \cite{HiraiMandel2022} and \cite{DiStefanoetal2022} were recently suggested, but are not yet implemented in the available codes. It would be interesting to compare the results presented in this manuscript with results obtained using different CEE formalisms.
 
The results of this study can be used to test whether NS/BH-core mergers and their resultant CEJSN transient events can account for several high energy astrophysical phenomena (as proposed and studied in e.g., \citealt{Papishetal2015}; \citealt{SokerGilkis2018}; \citealt{Sokeretal2019}; \citealt{GrichenerSoker2019a}; \citealt{GrichenerSoker2021}; \citealt{Soker2021}; \citealt{Gricheneretal2022}; \citealt{Soker2022b}) and what would be their overall contribution to said events. Mergers with different core structures, for instance, can lead to transient events with different properties. R-process nucleosynthesis in CEJSN requires that the merger of the giant's core with the NS occurs when the core is CO rich. \cite{GrichenerSoker2019a} found that one CEJSN r-process events per $\simeq 1000$ CCSNe suffices to explain the solar system r-process abundances by the CEJSN r-process scenario, which is consistent with our results for $\alpha_{\rm CE} \simeq 1.25$ (Fig. \ref{fig:CompactObjectCoreMergersRates} and the brown dots of Fig. \ref{fig:PercentHeCO}). For this scenario to account for a substantial fraction of the r-process abundance (above $10\%$)  $1.25 \lesssim \alpha_{\rm CE} \lesssim 1.75$ is required.  

\cite{GrichenerSoker2021} found that jets launched by a BH inside the envelope of a giant star might emit neutrinos with energies of $10^{15} \eV $ as detected by IceCube \citep{Aartsenetal2013}. The event rate required to explain the high-energy neutrino flux assuming the properties of the model described in \cite{GrichenerSoker2021}, and based on \cite{Aartsenetal2021} is $\approx 3 \%$ from CCSNe, and can be lower for jets with higher energies. We find that the percentage of systems that result in CEE of a BH with the giant secondary star (stage six in \textit{channel I}; left panel of Fig. \ref{fig:EvolutionRoutes}) is about $\simeq 0.6 \%$ from CCSNe for all values of $\alpha_{\rm CE}$ we simulate. This implies that the CEJSN scenario for high-energy neutrinos might have a significant contribution to the high-energy neutrino flux.

We note that even though the motivation for this work was the CEJSN scenario and its possible outcomes, the population model presented in this manuscript is not exclusive for the jets powering mechanism, and the merger rates shown in Fig. \ref{fig:CompactObjectCoreMergersRates} can be used for any model that aims to explain NS/BH-core mergers in both frameworks of traditional CEE ($\alpha_{\rm CE} \leq 1$), or models that assume another energy source besides the orbital energy ($\alpha_{\rm CE} > 1$). 

\section*{ACKNOWLEDGMENTS}
\label{sec:acknowledgments}

I thank Noam Soker, Vladimir Kalnitsky, Amit Kashi, Avishai Gilkis, Jan J. Eldridge, Alejandro Vigna-Gomez, Hila Glanz and Dmitry Shishkin for helpful discussions and important suggestions that helped in improving this manuscript. I thank an anonymous referee for detailed comments and suggestions that made this manuscript better. This research was supported by a grant from the Israel Science Foundation (769/20). I acknowledge support from Irwin and Joan Jacobs Fellowship. Simulations in this paper made use of the COMPAS rapid binary population synthesis code (version 02.31.06), which is freely available at http://github.com/TeamCOMPAS/COMPAS. 

\section*{Data availability}
\label{sec:DataAvailability}
The data underlying this article will be shared on reasonable request to the corresponding author.

\appendix 
\section*{The effect of mass accretion by the NS/BH during CEE with the giant star}
\label{sec:Macc}

The NS/BH merger with the core of the giant star can power a bright transient event. In the CEJSN scenario (e.g., \citealt{Sokeretal2019}) the NS/BH launches two opposite jets, one on each side of the equatorial plane, while spiralling-in inside the envelope and later through the core of the giant secondary star. The collision of these jets with the giant's envelope and the giant's winds converts kinetic energy to thermal energy and then radiation, leading to a bright transient that might mimic a SN explosion. 

The energy that the two jets a NS launches deposit to the envelope of the giant secondary star is \begin{equation}
E_{\rm 2j} = \eta \frac {G M_{\rm NS} M_{\rm acc,NS}}{R_{\rm NS}},
\label{eq:EjetsNS}
\end{equation}
where $M_{\rm NS}$ and $R_{\rm NS}$ are the mass and the radius of the NS, respectively, and $\eta $ is an efficiency parameter. Assuming that the energy of the jets is about the binding energy of the envelope $E_{\rm bind}$, as required to unbind the envelope considering that hydrodynamical simulations find that most of the envelope's mass remains bound at the end of CEE (e.g., \citealt{Passyetal2012}), we can find the amount of mass accreted by the NS in the CEE phase to be
\begin{equation}
\begin{split}
M_{\rm acc,NS}  \simeq 
 0.001 \, M_{\rm \odot}  \left(\frac{E_{\rm bind}}{5\times 10^{49} \erg} \right)\left(\frac{R_{\rm NS}}{10 \km} \right)  \\ \times \left(\frac{M_{\rm NS}}{1.25 \, M_{\rm \odot}}\right)^{-1}  \left(\frac{\eta}{0.05} \right)^{-1}.
\label{eq:MaccNS}
\end{split}
\end{equation}
We calibrate $E_{\rm bind}$, $R_{\rm NS}$ and $M_{\rm NS}$ with typical values from our \textsc{COMPAS} models and we take the efficiency parameter to be $\eta = 0.05$ (e.g., \citealt{LopezCamaraetal2019}). 

The jets that a BH launches inside the envelope of the giant secondary star are relativistic, and their energy is given by   
\begin{equation}
E_{\rm 2j} = \eta M_{\rm acc, BH} c^{2},
\label{eq:EjetsBH}
\end{equation}
implying that the amount of mass that the BH accretes while inside the envelope of the giant secondary star is
\begin{equation}\begin{split}
M_{\rm acc,BH}  \simeq 
 0.002 \, M_{\rm \odot}  \left(\frac{E_{\rm bind}}{3\times 10^{50} \erg} \right) \left(\frac{\eta}{0.1} \right)^{-1}.
\label{eq:MaccBH}
\end{split}
\end{equation}
We take the typical value of $E_{\rm bind}$ from the \textsc{COMPAS} simulations and the efficiency parameter $\eta = 0.1$ from \cite{Franketal2002}.

Including the mass that is accreted by the NS/BH during CEE in our \textsc{COMPAS} simulations yields practically no difference in the rates of NS/BH-core mergers for all values of $\alpha_{\rm CE}$ we simulated. The rates remain similar even for the higher bounds of the accreted masses inferred from substituting the upper limits of the parameters from the simulations in equations (\ref{eq:MaccNS}) and (\ref{eq:MaccBH}), which results in both cases in masses about an order of magnitude higher than the calibrated values. Inside the core of the giant star the energies of the jets are higher and hence the accretion has a more significant role.

\end{document}